\documentclass[a4paper,11pt]{article}
\usepackage{pos}
\usepackage{caption}
\usepackage{subcaption}
\usepackage{siunitx}
\usepackage{lineno}
\usepackage{wrapfig}
\usepackage{todonotes}

\author{The IceCube Collaboration \\{\normalsize \normalfont(a complete list of authors can be found at the end of the proceeding)}}

\emailAdd{lasse.halve@icecube.wisc.edu}
\emailAdd{johannes.werthebach@icecube.wisc.edu}

\abstract{The IceCube Upgrade is an extension of the IceCube detector at the geographic South Pole. It consists of seven new strings with novel instrumentation. More than 430 multi-PMT optical modules called "mDOMs", housing 24 3-inch PMTs each, will be produced for the Upgrade. This will require testing and pre-calibration on a short timescale of more than 10,000 PMTs prior to assembly and deployment. We present the design of a PMT testing facility that enables simultaneous testing of roughly 100 PMTs per day at temperatures down to -20°C. The design is implemented at RWTH Aachen University and TU Dortmund University in parallel to achieve a throughput of up to 1,000 PMTs per week. This will enable a steady supply of tested PMTs to the production sites, which is critical for the Upgrade, as well as the future IceCube-Gen2 project.

\vspace{4mm}
{\bfseries Corresponding authors:}
Lasse Halve$^{1*}$, Johannes Werthebach$^{2}$\\
{$^{1}$ \itshape RWTH Aachen University}\\
{$^{2}$ \itshape TU Dortmund University}\\[4mm]
$^*$ Presenter
}

\title{Design of an Efficient, High-Throughput Photomultiplier Tube Testing Facility for the IceCube Upgrade}
\ShortTitle{Photomultiplier Tests for the IceCube Upgrade mDOM}

\FullConference{37$^{\rm{th}}$ International Cosmic Ray Conference (ICRC 2021)\\
		July 12th -- 23rd, 2021\\
		Online -- Berlin, Germany}

\begin{document}
\maketitle

\section{Introduction}
\begin{figure}[htb]
    \centering
    \includegraphics[width=.45\textwidth]{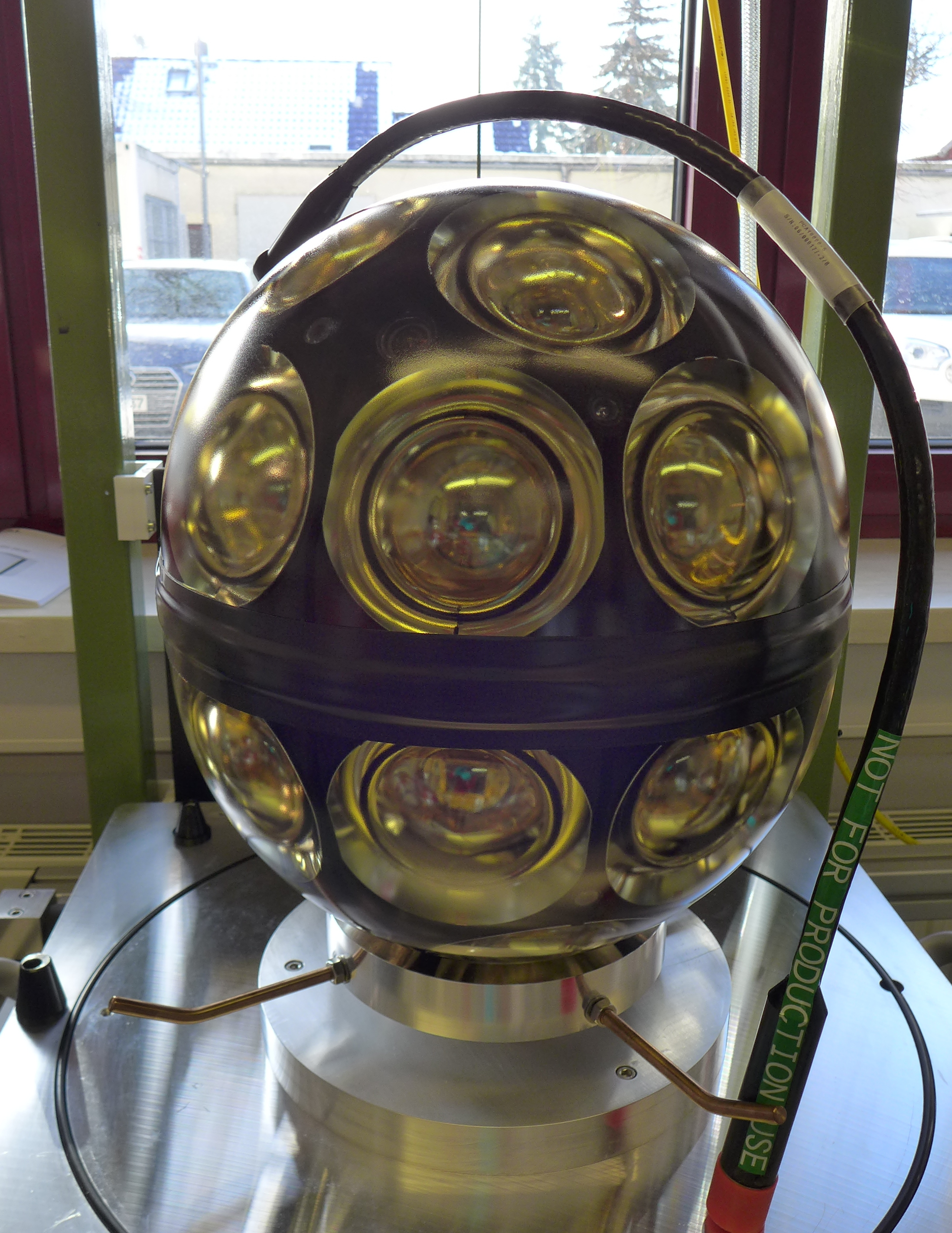}
    \caption{The first completed mDOM. Each hemisphere houses twelve PMTs as well as various calibration devices. Picture by Matthias Schust, DESY.}
    \label{fig:froggy}
\end{figure}

The IceCube Neutrino Observatory \cite{Aartsen:2016nxy} is the world’s largest neutrino telescope, and is located at the South Pole. IceCube consists of more than 5000 large area photomultipliers (PMT) inside Digital Optical Modules (DOM) attached to 86 cable-strings instrumenting roughly 1\,\si{\kilo\meter^3} of ice. These PMTs detect Cherenkov light emitted by charged particles produced in the interactions of neutrinos in the surrounding ice or nearby bedrock. IceCube was optimized to investigate high-energy neutrinos in the TeV to PeV energy scale. The DeepCore extension \cite{Abbasi:2012} lowered the energy threshold to $\sim$10\,GeV by a denser instrumented section in the center of the detector. This threshold will be further reduced by the upcoming IceCube Upgrade \cite{Ishihara:2019uL}. It will consist of seven cable-strings with up to 120 optical modules each, embedded near the bottom center of the existing IceCube Neutrino Observatory. Two new types of optical modules, the Multi-PMT Digital Optical Module (mDOM) \cite{mDOM} and the Dual optical sensors in an Ellipsoid Glass for Gen2 (D-Egg) \cite{DEgg}, are to be installed. The mDOM (see figure \ref{fig:froggy}) features a matrix of 24 3-inch photomultipliers (PMTs) \cite{mDOM_PMTs} as well as several calibration devices, such as fast LEDs \cite{Cal_LEDs}, CCD cameras \cite{Camera} and other on-board sensors. Since the PMTs are the primary detection unit for the mDOMs, they need to be tested for functionality before integration into  modules.

\section{Requirements of the PMT Testing Facility}
The main goal of the testing facility is ensuring the functionality of each PMT before it is installed in the sensor modules. Additionally, calibrations of required high voltage, photo-detection efficiency, charge response linearity, darkpulse rate, timing resolution, and probabilities of pre-, late-, and afterpulses will be done. These tests will be performed at a temperature of \SI{-20}{\celsius}, the typical ambient temperature of the deep South Pole ice \cite{Price7844}.

In total, over 10.000 PMTs have to be tested for the completion of all mDOM modules. In order to keep up with production timelines, the throughput of the testing facility has to be at a few hundred PMTs per week. This requires testing many PMTs at the same time as well as fast turnaround times between measurements. For the future, the adaptability of the testing facility to other types of PMTs is important as preparations for the construction of IceCube Gen2 \cite{Gen2_DOM} proceed.


\section{Mechanical Design and Implementation}
\begin{figure}[htbp]
    \centering
    \includegraphics[width=0.75\textwidth]{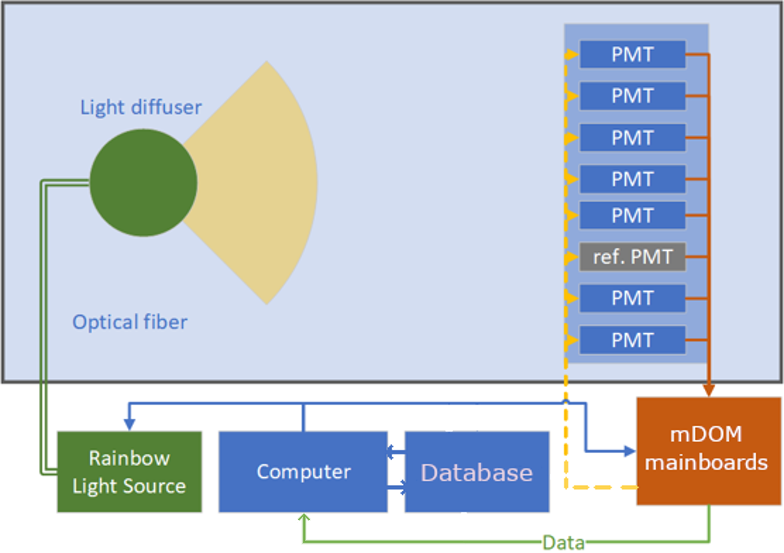}
    \caption{Schematic of the test facility design. The photomultipliers are mounted in a rack inside a cooling unit. They are supplied with high-voltage by active bases \cite{mDOM}. The readout and control of the bases and PMTs is done with mDOM mainboards \cite{mDOM}. Processing, storage, and analysis of data is performed by a computer running a local database. The computer controls all other hardware devices at the facility, e.g. the light source system. All electronics but the PMTs are kept outside the cooling unit to avoid temperature and humidity dependent effects. Light is routed into the cooling unit through an optical fiber that ends in a PTFE integration sphere from \cite{POCAM}, acting as a diffuser and illuminating the PMTs.}
    \label{fig:setup-schematic}
\end{figure}

\begin{figure}[htbp]
    \centering
    \begin{subfigure}[b]{0.45\textwidth}
        \centering
        \includegraphics[width=\textwidth]{figures/container.png}
        \caption{Shipping refrigeration container in the Aachen setup. It is equipped with additional insulation and cable entries. The container can cool down to \SI{-25}{\celsius}.}
        \label{fig:container}
    \end{subfigure}
    \hfill
    \begin{subfigure}[b]{0.45\textwidth}
        \centering
        \includegraphics[width=\textwidth]{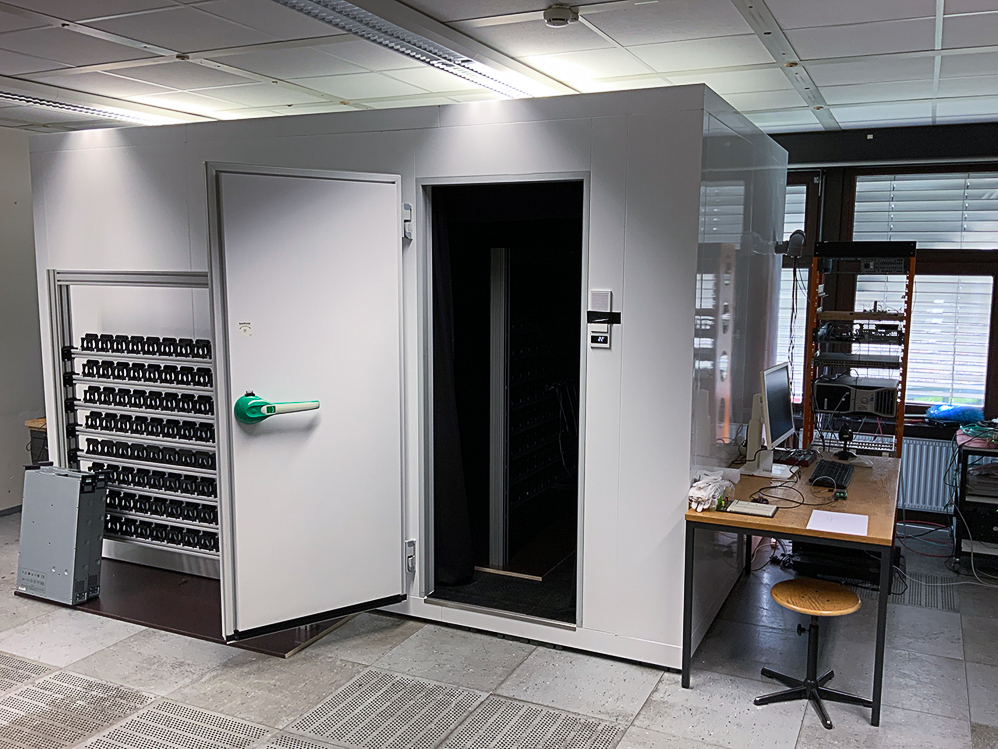}
        \caption{Off-the-shelf food industry freezing container in the Dortmund setup. It is modified with two cable entries and a light absorbing cloth on the inside.}
        \label{fig:cooling room}
    \end{subfigure}
    \caption{Solutions for climate controlled testing environments in Aachen and Dortmund.}
    \label{fig:cooling-units}
\end{figure}

PMT Testing facilities have been implemented at two sites: RWTH Aachen University and TU Dortmund University. Both follow the conceptual design seen in figure \ref{fig:setup-schematic}. A dark, temperature controlled room is needed for the tests. In Aachen, a commercial refrigeration container is used (see figure \ref{fig:container}). In Dortmund, a climate controlled chamber is used (see figure \ref{fig:cooling room}). Special care was taken to ensure that the rooms are completely light- and air-tight. Light leaking into the setup would lead to an enhanced darkpulse rate, air leaking in can cause problems of high air humidity, icing of components, and condensation during temperature cycles.

\begin{figure}[htbp]
    \centering
    \includegraphics[width=0.75\textwidth]{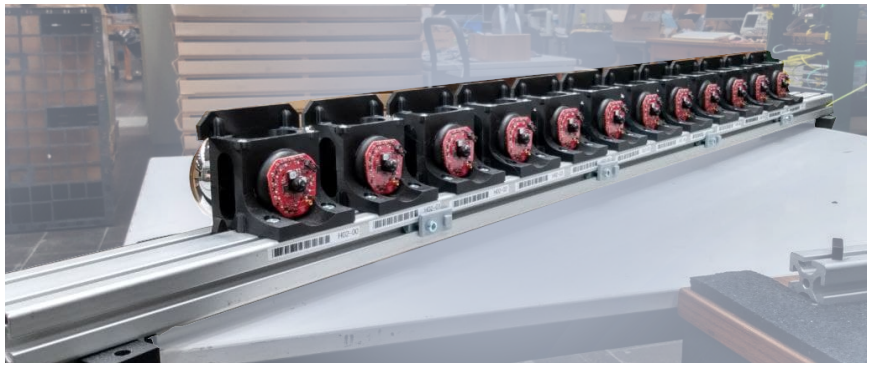}
    \caption{A slide-in bar with twelve mounted PMTs. The PMT mounts are 3D-printed and fastened on an aluminum extrusion profile. In red, the active bases \cite{mDOM} can be seen. Barcodes enable a direct association of PMT serial number and readout channel in the setup.}
    \label{fig:vogelstange}
\end{figure}

Inside the cooling rooms, a rack carrying eight slide-in  bars (see figure \ref{fig:vogelstange}), holding twelve PMTs each, is installed. In total, 96 PMTs can be installed in each rack at once. Using this modular approach, PMTs can be mounted outside the testing environment and quickly installed into the rack. PMTs, PMT holders, bars, as well as each rack position has unique barcode identifiers attached, that enable a direct association of the PMT serial number to the readout channel.

Each PMT connects to one of four mDOM mainboards \cite{mDOM} in an electronics rack outside the cooling room. Each mainboard is responsible for controlling and reading out 24 PMTs. This includes setting the high voltage for each PMT, recording waveforms, and reading scaler rates. The mainboards as well as all other hardware devices are controlled by a computer that also processes and stores data received from the mainboards.

Each testing cycle is composed of several measurements and analyses. The configurations for the testing is stored in a locally hosted database. The measured data is also stored in the database after processing, i.e. after baseline correction and feature extraction on waveforms. That data can instantaneously be accessed by fully automated analysis procedures that produce results, which are stored in a collaboration-wide database. This ensures that all testing and calibration data is well available and remains in permanent storage.


\section{Light Source System}

\begin{figure}[htbp]
    \centering
    \begin{subfigure}[b]{0.45\textwidth}
        \centering
        \includegraphics[width=\textwidth]{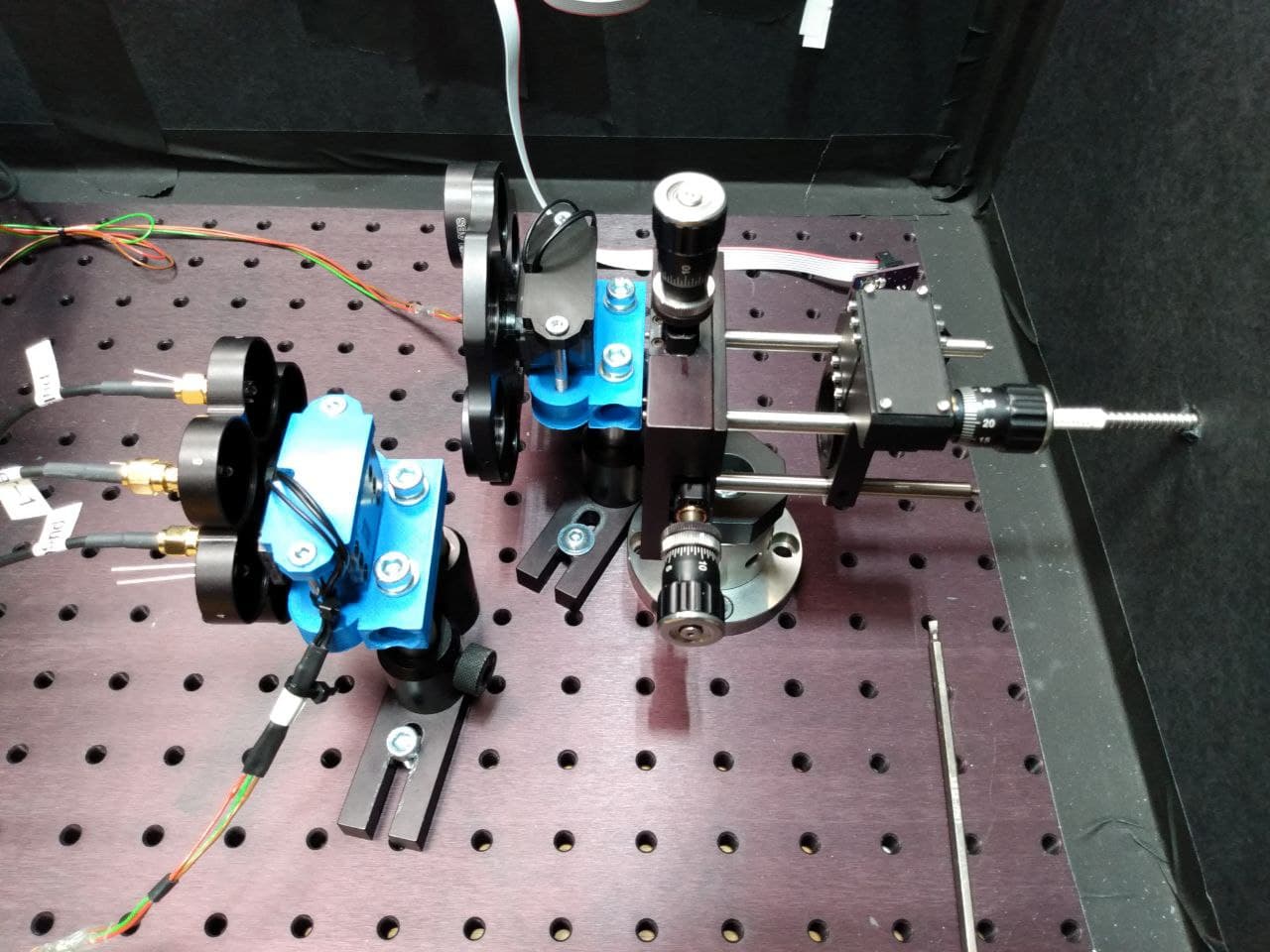}
        \caption{Light source system in the Aachen setup. LEDs driven by short electrical pulses \cite{mrongen} are coupled into an optical fiber. Single LEDs can be selected with a motor-driven wheel. Optionally, filters can be driven into the light's path.}
        \label{fig:thorlabs-lightsource}
    \end{subfigure}
    \hfill
    \begin{subfigure}[b]{0.45\textwidth}
        \centering
        \includegraphics[width=\textwidth]{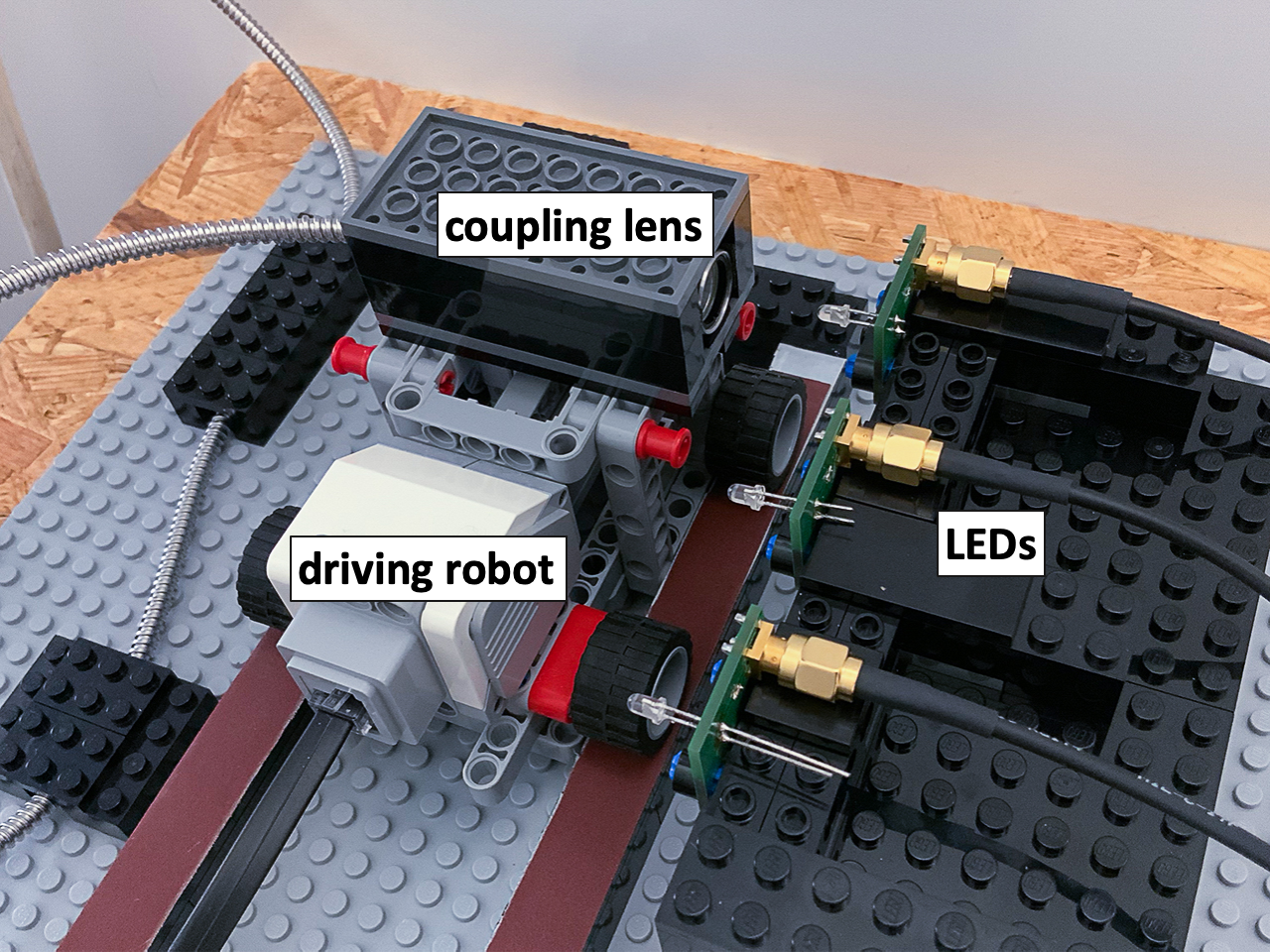}
        \caption{Light source system in the Dortmund setup. Single LEDs can be coupled into the fiber by driving the coupling lens in front of the LED. It is possible to install up to five different LEDs.\\}
        \label{fig:lego-lightsource}
    \end{subfigure}
    \caption{Light source solutions in Aachen and Dortmund.}
    \label{fig:lightsources}
\end{figure}

Testing the PMTs requires fast light sources of different wavelengths and intensities. Both facilities use the same design, but slightly different implementations (see figures \ref{fig:thorlabs-lightsource} and \ref{fig:lego-lightsource}). Outside the climate controlled room, LEDs (\SI{375}{\nano\meter}, \SI{505}{\nano\meter}) are mounted in a selection wheel, that is driven by a stepper motor. The LEDs are electrically driven by nanosecond-resolution pulsers \cite{mrongen}, which also provide a synchronous trigger signal to the mainboards. Light from the selected LED passes through an additional wheel with optional neutral-density filters and is coupled into an optical fiber. The fiber is routed inside the test setup and ends in an integrating PTFE sphere adapted from \cite{POCAM}, acting as a diffuser.

\begin{figure}
    \centering
    \includegraphics[width=0.75\textwidth]{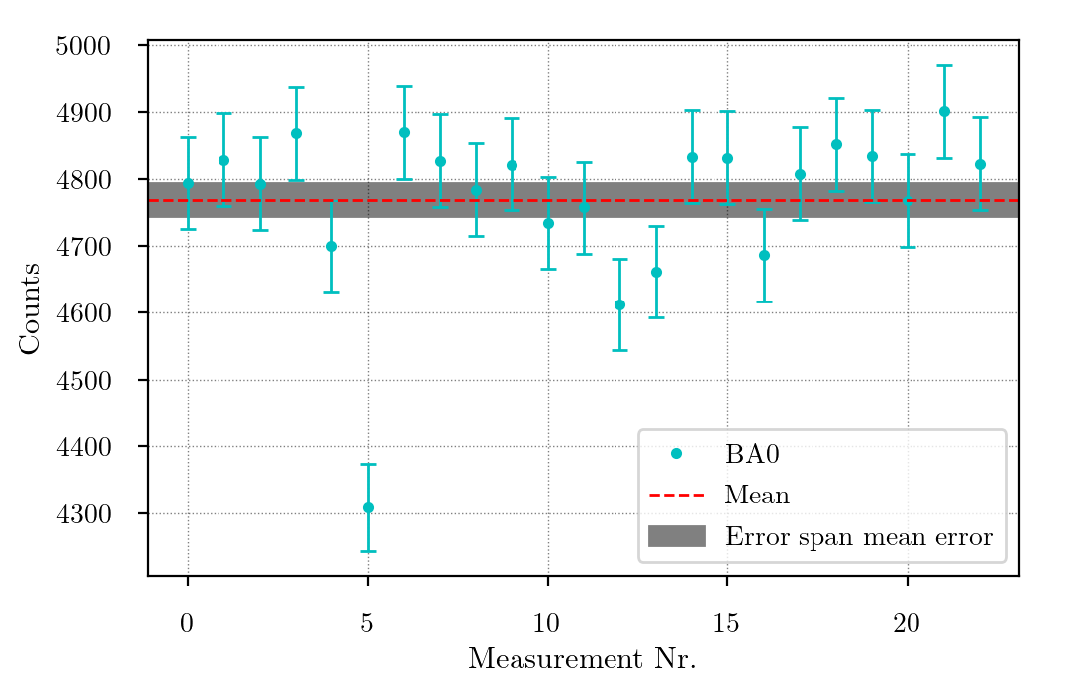}
    \caption{Light source brightness study. The teal markers show repeated rate measurements of the same PMT (BA0) after driving the light source selection wheel before each measurement. Shown are the number of detected pulses in 50,000 recorded waveforms triggered on the light source output. The deviation of the number of detected pulses is small compared to the total number of pulses ($\approx 1.5\%$). Notably, there is a large deviation in one measurement. This probably is an effect of non-optimal coupling of light into the optical fiber. Such deviations can later on be corrected for using well-calibrated reference PMTs.}
    \label{fig:lightsource-stability}
\end{figure}

To test the stability of the light yield, repeated measurements of the number of observed photons with a PMT have been performed \cite{Marco}. Between each measurement, the selection apparatus was driven away from and back to the LED in question. Figure \ref{fig:lightsource-stability} shows that there are a few measurements with decreased brightness. In measurements sensitive to the photon yield, this can be corrected for by including known reference PMTs into the setup.

\begin{figure}[htb!]
    \centering
    \includegraphics[width=0.75\textwidth]{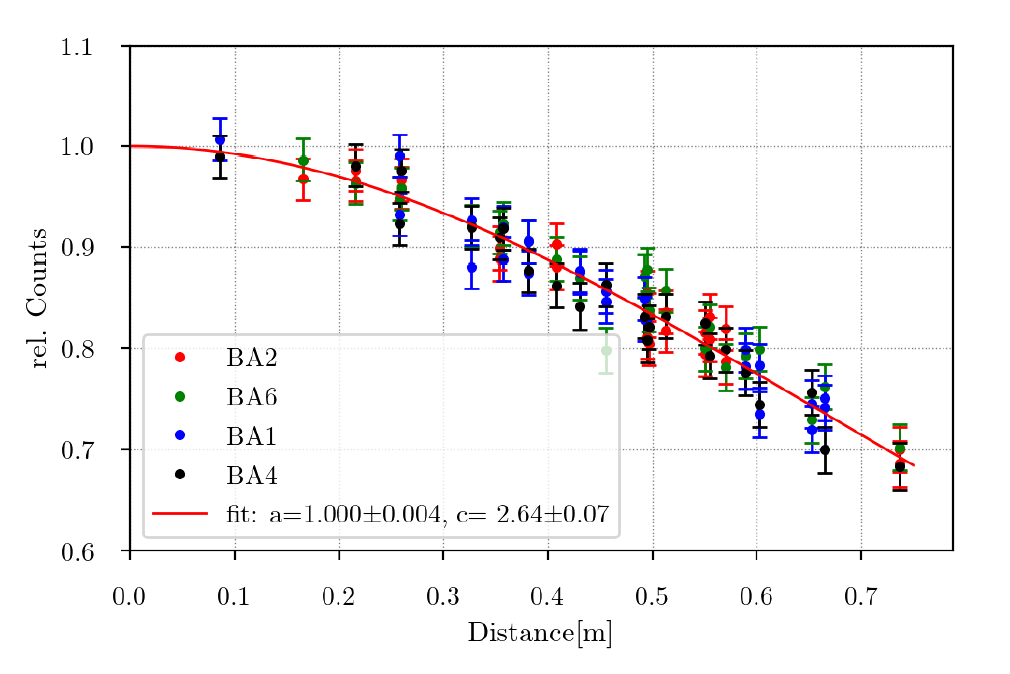}
    \caption{Flat field correction calibration. Each set of colored markers shows measurements performed with a different PMT (BAx) in multiple positions in the PMT rack. The number of detected pulses relative to the mean number of detected pulses of two reference PMTs in the center of the rack is shown as a function of the distance from the center of the grid. The number of detected pulses has been corrected for the individual relative photo-detection efficiencies of the PMTs. The red line shows a combined fit of equation \ref{eq:flat-field-correction} to the data.}
    \label{fig:flat-field-correction}
\end{figure}

For measurements of the relative photo-detection efficiency, the relative amount of light at each position of the rack has to be known and corrected for. With perfectly isotropic light, we expect a quadratic dependency of the light yield as a function of the distance to the center of the rack. Additionally, the effective area of the PMTs decreases with larger angles between the PMT's symmetry axis and the incident light. In total, the relative light yield can be described by equation \ref{eq:flat-field-correction}, where $C_\mathrm{rel.}$ is the relative light yield, $R_0$ is the distance from the diffuser to the center of the PMT rack, $x$ is the distance of the PMT position to the center of the rack, and $a$ and $c$ are free parameters
\begin{equation}
    C_{\mathrm{rel.}} = a \cdot \left({\frac{R_0}{\sqrt{R_0^2 + x^2}}}\right)^c.\\\label{eq:flat-field-correction}
\end{equation}
A calibration measurement of this dependency is shown in figure \ref{fig:flat-field-correction}. Four PMTs moving around the rack were used to cover all positions in the rack. The exponent $c$ in the combined fit yields $2.64\pm0.07$. This parametrization will be used for the flat field correction in photo-detection efficiency measurements.

\section{Results from Test Runs}
\begin{figure}[htbp!]
    \centering
    \includegraphics[width=0.8\textwidth]{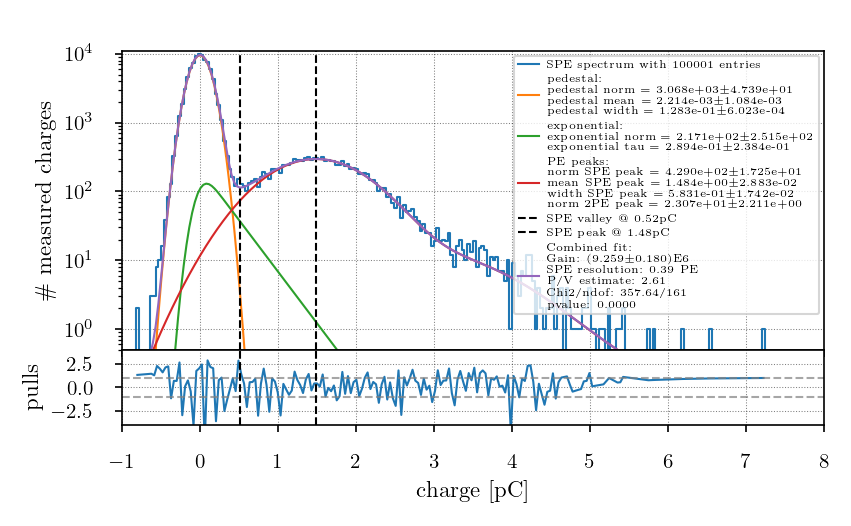}
    \caption{Single-Photo-Electron spectrum of PMT DM00168 at voltages of \SI{96}{\volt} between dynodes. The blue histogram shows the measured charges with an external trigger from the light source system. The colored lines show a fit to the histogram with three components: The orange line is a Gaussian describing the pedestal region, the red line is a combination of Gaussians for the 1\,PE and 2\,PE peak, the green line is an exponential with cutoff at zero describing badly amplified photo-electrons. One can directly read off the gain of the PMT from the position of the SPE peak at a value of \SI{1.48}{\pico\coulomb}.}
    \label{fig:example-spe-plot}
\end{figure}

For the construction of the first ten modules, several PMTs have undergone testing with a reduced set of measurements and analyses. Fully automated measurements of Single-Photo-Electron (SPE) spectra and calibration of target high voltage are implemented. Figure \ref{fig:example-spe-plot} shows such a recorded SPE spectrum. The recorded SPE spectrum can be well described by a simple fit function (see caption). From the fit results, the gain and other characteristics of the PMT are extracted. In total, only a single PMT out of the first 320 was rejected because of a faulty solder joint.


\clearpage
\section{Summary and Outlook}
The construction of mDOMs for the IceCube Upgrade requires fast testing of many PMTs. A design for a facility testing up to 96 PMTs simultaneously is implemented at two sites: RWTH Aachen University and TU Dortmund University. Custom light source systems have been constructed and tested at both sites. A modular mounting system of slide-in bars reduces the time between testing cycles. The testing procedures, data processing, analysis, and storage have been fully automized. The mounts can easily be adapted to fit PMTs of other sizes, such that the design can also be used for future IceCube-Gen2 testing. In the near future, these testing facilities will be used to test every single PMT that will be integrated into the mDOM modules.


\bibliographystyle{ICRC}
\bibliography{references}

\clearpage
\section*{Full Author List: IceCube Collaboration}




\scriptsize
\noindent
R. Abbasi$^{17}$,
M. Ackermann$^{59}$,
J. Adams$^{18}$,
J. A. Aguilar$^{12}$,
M. Ahlers$^{22}$,
M. Ahrens$^{50}$,
C. Alispach$^{28}$,
A. A. Alves Jr.$^{31}$,
N. M. Amin$^{42}$,
R. An$^{14}$
% ...
\\

\noindent
% ...
$^{12}$ Universit{\'e} Libre de Bruxelles, Science Faculty CP230, B-1050 Brussels, Belgium \\
% ...
$^{14}$ Department of Physics and Laboratory for Particle Physics and Cosmology, Harvard University, Cambridge, MA 02138, USA \\
% ...
$^{17}$ Department of Physics, Loyola University Chicago, Chicago, IL 60660, USA \\
$^{18}$ Dept. of Physics and Astronomy, University of Canterbury, Private Bag 4800, Christchurch, New Zealand \\
% ...
$^{22}$ Niels Bohr Institute, University of Copenhagen, DK-2100 Copenhagen, Denmark \\
% ...
$^{28}$ D{\'e}partement de physique nucl{\'e}aire et corpusculaire, Universit{\'e} de Gen{\`e}ve, CH-1211 Gen{\`e}ve, Switzerland \\
% ...
$^{31}$ Karlsruhe Institute of Technology, Institute for Astroparticle Physics, D-76021 Karlsruhe, Germany  \\
% ...
$^{42}$ Bartol Research Institute and Dept. of Physics and Astronomy, University of Delaware, Newark, DE 19716, USA \\
% ...
$^{50}$ Oskar Klein Centre and Dept. of Physics, Stockholm University, SE-10691 Stockholm, Sweden \\
% ...
$^{59}$ DESY, D-15738 Zeuthen, Germany \\
% ...

\subsection*{Acknowledgements}

\noindent
USA {\textendash} U.S. National Science Foundation-Office of Polar Programs,
U.S. National Science Foundation-Physics Division,
U.S. National Science Foundation-EPSCoR,
Wisconsin Alumni Research Foundation,
Center for High Throughput Computing (CHTC) at the University of Wisconsin{\textendash}Madison,
Open Science Grid (OSG),
Extreme Science and Engineering Discovery Environment (XSEDE),
Frontera computing project at the Texas Advanced Computing Center,
U.S. Department of Energy-National Energy Research Scientific Computing Center,
Particle astrophysics research computing center at the University of Maryland,
Institute for Cyber-Enabled Research at Michigan State University,
and Astroparticle physics computational facility at Marquette University;
Belgium {\textendash} Funds for Scientific Research (FRS-FNRS and FWO),
FWO Odysseus and Big Science programmes,
and Belgian Federal Science Policy Office (Belspo);
Germany {\textendash} Bundesministerium f{\"u}r Bildung und Forschung (BMBF),
Deutsche Forschungsgemeinschaft (DFG),
Helmholtz Alliance for Astroparticle Physics (HAP),
Initiative and Networking Fund of the Helmholtz Association,
Deutsches Elektronen Synchrotron (DESY),
and High Performance Computing cluster of the RWTH Aachen;
Sweden {\textendash} Swedish Research Council,
Swedish Polar Research Secretariat,
Swedish National Infrastructure for Computing (SNIC),
and Knut and Alice Wallenberg Foundation;
Australia {\textendash} Australian Research Council;
Canada {\textendash} Natural Sciences and Engineering Research Council of Canada,
Calcul Qu{\'e}bec, Compute Ontario, Canada Foundation for Innovation, WestGrid, and Compute Canada;
Denmark {\textendash} Villum Fonden and Carlsberg Foundation;
New Zealand {\textendash} Marsden Fund;
Japan {\textendash} Japan Society for Promotion of Science (JSPS)
and Institute for Global Prominent Research (IGPR) of Chiba University;
Korea {\textendash} National Research Foundation of Korea (NRF);
Switzerland {\textendash} Swiss National Science Foundation (SNSF);
United Kingdom {\textendash} Department of Physics, University of Oxford.

\end{document}